\documentclass[twocolumn,showpacs,superscriptaddress,prb]{revtex4-1}
\usepackage{graphicx}
\usepackage{color}
\usepackage{amsmath,amsthm,amssymb}
\usepackage{braket}
\usepackage[colorlinks,citecolor=blue,linkcolor=red]{hyperref}

\usepackage{soul}

\begin{document}

\title{Landscape of coexisting excitonic states\\in the insulating single-layer cuprates and nickelates}

\author{Christopher Lane}
\email{laneca@lanl.gov}
\affiliation{Theoretical Division, Los Alamos National Laboratory, Los Alamos, New Mexico 87545, USA}
\affiliation{Center for Integrated Nanotechnologies, Los Alamos National Laboratory, Los Alamos, New Mexico 87545, USA}

\author{Jian-Xin Zhu}
\email{jxzhu@lanl.gov}
\affiliation{Theoretical Division, Los Alamos National Laboratory, Los Alamos, New Mexico 87545, USA}
\affiliation{Center for Integrated Nanotechnologies, Los Alamos National Laboratory, Los Alamos, New Mexico 87545, USA}

\date{\today} 
\begin{abstract}
We present an {\it ab initio} study of the excitonic states of a prototypical high-temperature superconductor La$_2$CuO$_4$ and compare them to the isostructural single-layer nickelate La$_2$NiO$_4$. Key difference in the low-energy electronic structure leads to very different excitonic behavior. Excitons in La$_2$CuO$_4$ are delocalized and can freely move in the CuO$_2$ plane without disturbing the antiferromagnetic order. In contrast, in La$_2$NiO$_4$ we find the low-lying excitonic states to be extremely localized, producing a nearly flat dispersion. The theoretically obtained excitonic dispersion and behavior are in excellent agreement with RIXS observations. To classify the excitons we project the electron-hole coupling onto each atomic site including the full manifold of atomic orbitals. We find the excitons to be composed of a linear combination of exciton classes, including Mott-Hubbard, $d-d$, and charge-transfer.  The implication of these excitations to the high-$T_c$ pairing mechanism is also discussed.
\end{abstract}


\maketitle

\section{Introduction}
A long standing problem in the phenomenology of strongly correlated transition-metal oxides (TMO) is the nature of the insulator-metal transition and its close connection to the character of the electronic band gap.  The parent insulating phase typically falls into one of two categories: Mott or charge-transfer. In Mott insulators, the band gap is formed by the upper and lower Hubbard bands, since the on-site potential $U$ is less than the charge-transfer energy characterized by the ligand oxygen $2p$ levels. In contrast, the electronic gap in a charge transfer insulator is formed by oxygen $2p$ states and the upper Hubbard band. Physically, these two cases give rise to very different scenarios upon carrier doping; one where the carrier sits on the  transition-metal atom or the other where it sites on the oxygen sites. 

The classification of the gap alone, however, does not dictate the overall nature of doped state. For example, La$_2$CuO$_4$ (LCO) and La$_2$NiO$_4$ (LNO) are isostructural Ruddlesden-Popper transition-metal perovskites generally regarded as charge-transfer insulators, but they exhibit wildly different properties. LCO is a prototypical high-temperature superconductor with active itinerant carriers setting in around $x\approx 0.05$,~\cite{kastner1998magnetic} while no superconductivity has been reported in LNO with metallic behavior arising near $x\approx 0.8$.~\cite{eisaki1992electronic} The relationship between these two materials is made even more intriguing by the recent discovery of superconductivity in the infinite-layer nickelate.~\cite{li2019superconductivity}

To capture the nature of the occupied and unoccupied states, recent advances in high resolution resonant inelastic x-ray scattering (RIXS) provide a window into the energy and dispersion of elementary electronic excitations. Therefore, it provides a means to confront the low-energy elementary electronic excitations with theoretical predictions of electronic structure and dynamics. Of particular interest, RIXS measurements reveal opposing behaviors in the electronic excitation spectrum of LCO and LNO. In LCO, low energy excitonic bound states are found to be highly mobile displaying a parabolic energy dispersion from Brillouin zone center to zone edge.~\cite{collart2006localized,kim2002resonant} However, LNO exhibits completely localized electron-hole pairs, showing no energy dispersion as a function of momentum.~\cite{collart2006localized} 

An accurate first-principles treatment of the ground state electronic and magnetic structure of correlated materials is a fundamental challenge, and predicting emergent excited states further increases the complexity. The complete failure of the local density and generalized gradient approximations within the Hohenberg-Kohn-Sham\cite{hohenberg1964inhomogeneous,Kohn1965} density functional theory (DFT) in La$_2$CuO$_4$ ushered in the common belief that the density functional theory framework was fundamentally limited. Out of this void, many `beyond' DFT treatments, such as DFT+$U$,~\cite{dudarev1998electron,liechtenstein1995density,pesant2011dft+,czyzyk1994local} quasiparticle GW,~\cite{das2014intermediate} and various dynamical mean-field theory (DMFT) based schemes~\cite{held2006realistic,park2008cluster,kotliar2006electronic} were constructed to rationalize the low-energy spectra of La$_2$CuO$_4$ and many other correlated materials. However, these methodologies introduce external parameters, such as the on-site Hubbard $U$, to tune the correlation strength, which fundamentally  limit the predictive power. 

The theoretical investigation of excitons in strongly correlated matter has a long history starting around the time of BCS theory. Questions regarding exciton condensation,~\cite{blatt1962bose,keldysh1968collective,rademaker2013exciton,imada2019excitons,markiewicz2017excitonic,montiel2017effective,montiel2017local} propagation,~\cite{doniach1971excitons,moriya1972excitons} electron-hole pairing pathways,~\cite{zhang1998theory,barford2002excitons,matiks2009exciton,wrobel2002excitons,pouchard2008real,simon1996excitons} and their possible link to the mechanism of high-termperature superconductivity~\cite{ginzburg1970excitonic,allender1973model,weber1988cud,weber1989cu,jarrell1988charge,imada2019excitons} have been pursued. In particular, several calculations have been put forth classifying the excitons as charge-transfer, where the electron and hole site on neighboring Cu and O sites, along with justifying their dispersive nature 
in La$_2$CuO$_4$.~\cite{zhang1998theory,barford2002excitons,wrobel2002excitons} Additionally, Mott and $d-d$ excitons, where electron and holes originate from the Cu-$d_{x^2-y^2}$ bands and the $d_{x^2-y^2}/d_{z^2}$ orbitals, respectively, have also been suggested.\cite{markiewicz2017excitonic} However, a detailed characterization of the dispersion and nature of the excitons -- Mott vs. $d$-$d$ vs. charge-transfer --  in the real materials requires approaches that are not restricted to simple bases-sets and limiting cases.

Recent progress in constructing advanced density-functionals presents a new path forward in addressing the electronic structures of correlated materials at the first-principles level. In particular, the strongly-constrained-and-appropriately-normed (SCAN) meta-GGA exchange-correlation functional,~\cite{Sun2015}  has  been used to accurately predict many key properties of the undoped and doped La$_2$CuO$_4$ and YBa$_2$Cu$_3$O$_6$.~\cite{lane2018antiferromagnetic,furness2018accurate,zhang2020competing} In La$_2$CuO$_4$, SCAN correctly captures the magnetic moment in magnitude and orientation, the magnetic exchange coupling parameter, and the magnetic form factor along with the electronic band gap, all in accord with the corresponding experimental values. Recently, by treating the charge, spin, and lattice degrees of freedom on the same footing in a fully self-consistent manner the SCAN functional stabilizes 26 competing uniform and stripe phases in near-optimally doped YBa$_2$Cu$_3$O$_7$ without invoking any 
free parameters.~\cite{zhang2020competing} These results indicate that SCAN correctly captures many key features of the electronic and magnetic structures of the cuprates and thus provides a next-generation standard for investigating missing correlation effects.~\cite{lane2019iridate} We note that the transferability of SCAN to the wider class of transition-metal oxides has been demonstrated in Refs.~\onlinecite{varignon2019mott,zhang2019symmetry}.

In this article, we show that the excitonic dispersion in La$_2$CuO$_4$ and La$_2$NiO$_4$ can be captured within the DFT framework. Our first-principles, parameter-free magnetic ground state reproduces the key experimentally observed excitonic properties of La$_2$CuO$_4$ and La$_2$NiO$_4$. By projecting the electron-hole coupling matrix on to atomic-sites using the full manifold of atomic orbitals, we find the excitons to be composed of a linear combination of states, including Mott-Hubbard, $d-d$, and charge-transfer. Furthermore, we comment on the role these excitations may play in the superconducting pairing mechanism.

\section{Computational Methodology}

\begin{figure*}[ht]
\includegraphics[width=0.99\textwidth]{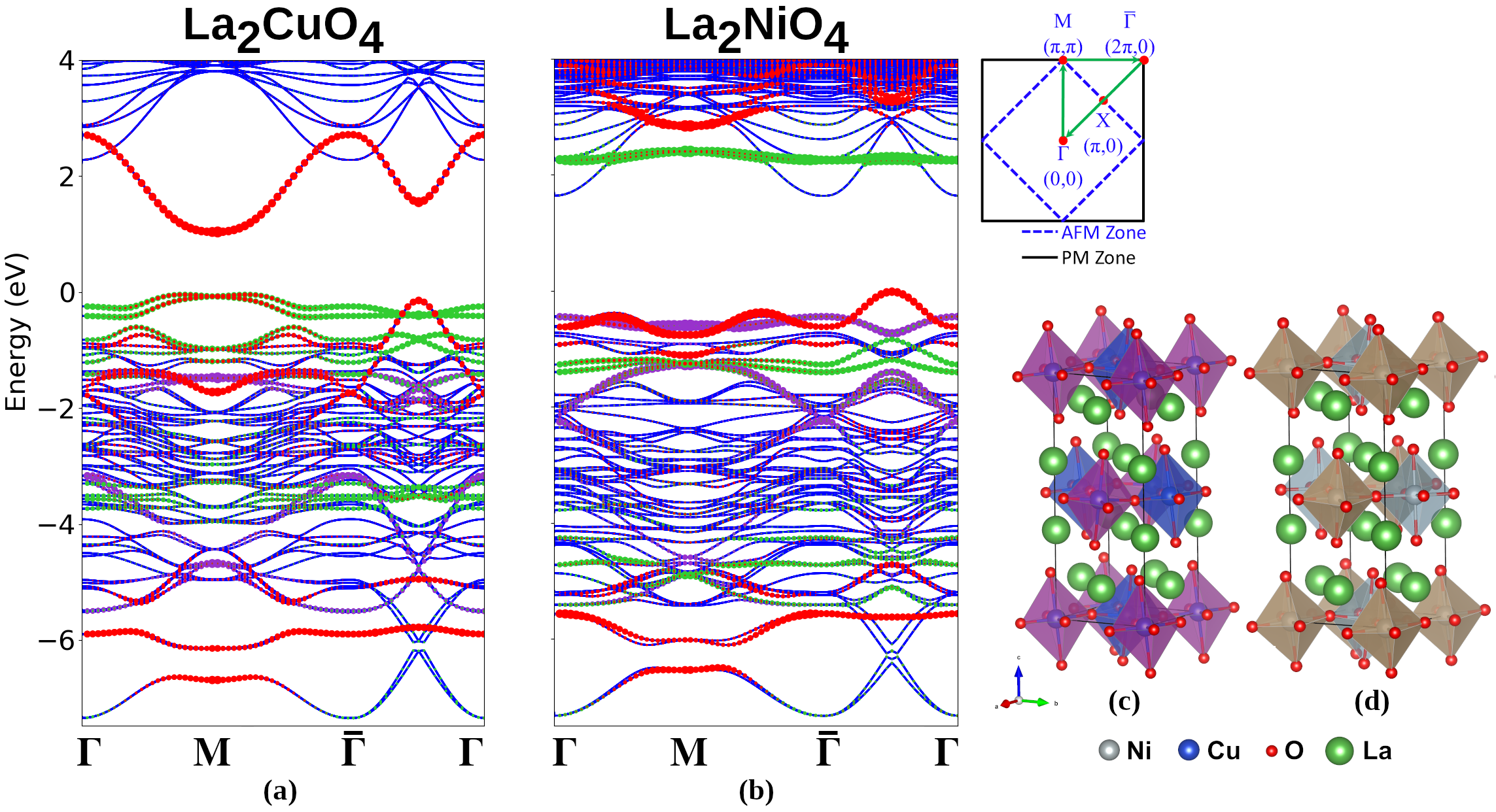}
\caption{(color online) (a) and (b) Electronic band structures (blue lines) along the high-symmetry lines in the Brillouin zone for La$_2$CuO$_4$ and La$_2$NiO$_4$ in the AFM phase. Contribution of $d_{x^2-y^2}$, $d_{z^2}$, and $d_{xy}$  orbitals are highlighted with red, green, and purple dots, respectively. The sizes of the dots are proportional to the fractional weights of orbital species. A schematic diagram of the AFM and reference tetragonal Brillouin zones with the path followed in presenting the band structures is shown on the right.   (c) and (d) Theoretically predicted AFM state of La$_2$CuO$_4$ and La$_2$NiO$_4$ in the  LTO and  LTT crystal structure, respectively. The related AFM structure is highlighted by coloring the octahedra in La$_2$CuO$_4$ blue (pink) and in La$_2$NiO$_4$ silver (brown) for spin-up (down). The in-plane oxygen atoms have no net magnetic moment for either compound. The black lines mark the unit cell.} 
\label{fig:BANDSanndSTRUCTURE}
\end{figure*}

{\it Ab initio} calculations were carried out by using the pseudopotential projector-augmented wave (PAW) method~\cite{Kresse1999} implemented in the Vienna ab initio simulation package (VASP)~\cite{Kresse1996,Kresse1993} with an energy cutoff of $500$ eV for the plane-wave basis set. The GW PAW potentials released with VASP.5.4 were used. Exchange-correlation effects were treated using the SCAN meta-GGA scheme.\cite{Sun2015} A 9 $\times$ 9 $\times$ 1 $\Gamma$-centered k-point mesh was used to sample the Brillouin zone. For La$_2$CuO$_4$ and La$_2$NiO$_4$ we used the low-temperature orthorhombic (LTO) and low-temperature tetragonal (LTT)  crystal structure of $Bmab$ and $P4_2/ncm$ symmetry, respectively, in accord with the experimentally observed structures.\cite{jorgensen1988superconducting,rodriguez1991neutron} All sites in the unit cell along with the unit cell dimensions were relaxed using a conjugate gradient algorithm to minimize energy with an atomic force tolerance of 0.008 eV/\AA. A total energy tolerance of $10^{-6}$ eV was used to determine the self-consistent charge density. The theoretically obtained structural parameters are in good accord with the corresponding experimental results.  As shown in Fig.~\ref{fig:BANDSanndSTRUCTURE}(c), the LTO structure can be viewed as being a $\sqrt{2} \times \sqrt{2}$ body-centered-tetragonal superlattice of I4/mmm symmetry in which $a^{\prime} \approx b^{\prime} \approx \sqrt{2}a$; the CuO$_6$ octahedra are rotated along the $(110)$ and $(1\bar{1}0)$ directions in alternate layers. The LTT structure (Fig.~\ref{fig:BANDSanndSTRUCTURE}(d)) is similar to the LTO structure, except $a^{\prime} = b^{\prime} = \sqrt{2}a$ and the NiO$_6$ octahedra are rotated along the $(100)$ and $(010)$ directions in alternate layers. 

The response functions and exciton eigenvalue calculations were carried out using the  screened interaction W and Bethe-Salpeter equation (BSE) as implemented in VASP. Following Liu {\it et al.}, we adopted the single-shot $W_0$ variant of the fully self-consistent screened interaction commonly employed in the $GW$ approximation due to its reasonable computational performance, while maintaining robust results.~\cite{liu2018relativistic} For the calculation of the response functions at the $W_0$ level, 125 frequency points and 600 virtual orbitals were used with an energy cutoff equal to half of the plane-wave cutoff. Erg{\"o}nenc {\it et al.}~\cite{ergonenc2018converged} demonstrated by a systematic analysis of the convergence of $G_0W_0$ results for a representative dataset of 3$d$, 4$d$, and 5$d$ TMO perovskites that 600 virtual orbitals are sufficient to obtain well converged results.

Since we wish to obtain excitons of zero and finite center-of-mass momentum, $\mathbf{Q}$, the BSE was solved beyond the Tamm-Dancoff approximation (TDA), including resonant-antiresonant coupling,
\begin{subequations}
\begin{align}
( E_{c\mathbf{k}+\mathbf{Q}}-E_{v\mathbf{k}} )&A_{vc\mathbf{k}}^{S,\mathbf{Q}}
+
\sum_{v^{\prime}c^{\prime}\mathbf{k}^{\prime}} K^{AA}_{vc\mathbf{k},v^{\prime}c^{\prime}\mathbf{k}^{\prime}}(\mathbf{Q})A_{v^{\prime}c^{\prime}\mathbf{k}^{\prime}}^{S,\mathbf{Q}}
\\
+
&\sum_{v^{\prime}c^{\prime}\mathbf{k}^{\prime}} K^{AB}_{vc\mathbf{k},v^{\prime}c^{\prime}\mathbf{k}^{\prime}}(\mathbf{Q})B_{v^{\prime}c^{\prime}\mathbf{k}^{\prime}}^{S,\mathbf{Q}}
=
\Omega^{S,\mathbf{Q}}A_{vc\mathbf{k}}^{S,\mathbf{Q}},\nonumber
\\
( E_{c\mathbf{k}+\mathbf{Q}}-E_{v\mathbf{k}} )&B_{vc\mathbf{k}}^{S,\mathbf{Q}}
+
\sum_{v^{\prime}c^{\prime}\mathbf{k}^{\prime}} K^{BA}_{vc\mathbf{k},v^{\prime}c^{\prime}\mathbf{k}^{\prime}}(\mathbf{Q})A_{v^{\prime}c^{\prime}\mathbf{k}^{\prime}}^{S,\mathbf{Q}}
\\
+
&\sum_{v^{\prime}c^{\prime}\mathbf{k}^{\prime}} K^{BB}_{vc\mathbf{k},v^{\prime}c^{\prime}\mathbf{k}^{\prime}}(\mathbf{Q})B_{v^{\prime}c^{\prime}\mathbf{k}^{\prime}}^{S,\mathbf{Q}}
=
-\Omega^{S,\mathbf{Q}}B_{vc\mathbf{k}}^{S,\mathbf{Q}},\nonumber
\end{align}
\end{subequations}
where $\Omega^{S,\mathbf{Q}}$ is the $S^{th}$ excitonic energy with center-of-mass momentum $\mathbf{Q}$ and $A_{vc\mathbf{k}}^{S,\mathbf{Q}}$ $(B_{vc\mathbf{k}}^{S,\mathbf{Q}})$ is the resonant (antiresonant) electron-hole coupling coefficient. Moreover, $W_0$ was used as a starting point for the construction of the screening properties in the interaction kernel $K_{vc\mathbf{k},v^{\prime}c^{\prime}\mathbf{k}^{\prime}}$. \footnote{It has also been shown that for finite-momentum the TDA breaks down for nanoscale systems \cite{ma2009excited,puschnig2013excited,gruning2009exciton} and to deviate from experiments in silicon.\cite{sander2015beyond}} 

\begin{figure*}[ht]
\includegraphics[width=0.99\textwidth]{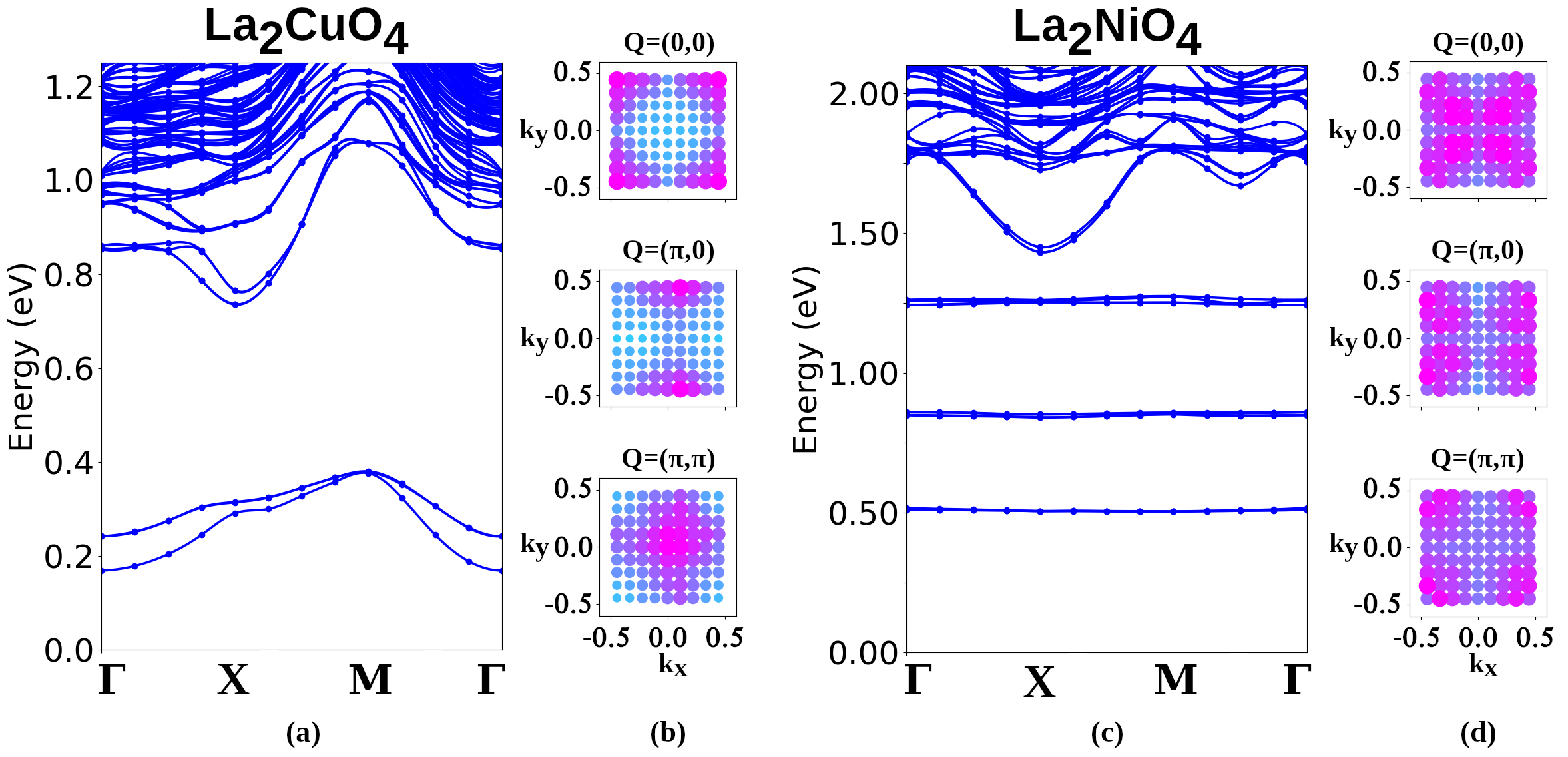}
\caption{(color online) (a) and (c) excitonic dispersion (blue dots and lines) along the high-symmetry lines in the Brillouin zone of momentum transfer for La$_2$CuO$_4$ and La$_2$NiO$_4$ in the AFM phase. (b) and (d) The contribution of each momenta $\mathbf{k}$ in first Brillouin zone to the energetically lowest exciton at each high-symmetry $\mathbf{Q}$-point. See text for details.} 
\label{fig:EXCBANDS}
\end{figure*}

SCAN is constructed within the generalized Kohn-Sham (gKS) scheme~\cite{seidl1996generalized} where the exchange-correlation potential is formally constructed to be orbital dependent and thus is ``non-multiplicative," in contrast to the ``multiplicative" potentials constructed within the LDA and GGA KS approaches. As a consequence, the gKS band gap is equal to the fundamental band gap in the solid, which is defined as the ground-state energy difference between systems with different numbers of 
electrons.~\cite{perdew2017understanding} In line with this result, recent SCAN-based studies obtain band gaps in the high-temperature cuprates and $3d$ transition-metal perovskite oxides in accord with experimental observations.~\cite{lane2018antiferromagnetic,furness2018accurate,varignon2019origin,lane2019iridate} 
This enables us to avoid the GW quasiparticle corrections and use directly the generalized Kohn-Sham band energies as the eigenvalues of the electrons $(E_{c\mathbf{k}+\mathbf{Q}})$ and holes $(E_{v\mathbf{k}})$ in the BSE Hamiltonian, where only seven conduction and  seven valence bands were considered.

\section{Ground State electronic and magnetic structure}

Figure~\ref{fig:BANDSanndSTRUCTURE}(a) and \ref{fig:BANDSanndSTRUCTURE}(c) show the electronic band structure (blue lines) and crystal structure of La$_2$CuO$_4$ in the  LTO structure for the antiferromagnetic (AFM) phase. The site-resolved atomic projections for $d_{x^2-y^2}$ (red dots), $d_{z^2}$ (green dots), and $d_{xy}$ (violet dots) are overlaid. Since the copper atoms have an oxidation state of $2^{+}$, only the $d_{x^2-y^2}$ is half-filed. Due to an intermediate electron-electron interaction ($U\approx 4.846$ eV)\cite{lane2018antiferromagnetic},  a moment of $0.487$ $\mu_{B}$ is produced in the $d_{x^2-y^2}$ orbital, with very little contribution from the rest of the $d$ manifold-of-states.~\footnote{Since we use a 2D k-point mesh to sample the brillouin zone the copper magnetic moment is slightly reduced compared to the results obtained with SCAN in Ref.~\onlinecite{lane2018antiferromagnetic}.} As a result of the AFM order, a $1.0$ eV band gap is formed in the  $d_{x^2-y^2}$ dominated band, with concomitant splitting around $-7$ eV. For more details and a thorough study of the ground state magnetic and electronic structure of La$_2$CuO$_4$ employing the SCAN functional please refer to Refs.~\onlinecite{lane2018antiferromagnetic,furness2018accurate}.

The magnetic and electronic structure La$_2$NiO$_4$ [Fig.~\ref{fig:BANDSanndSTRUCTURE}(b) and(d)] is similar to that of La$_2$CuO$_4$, except for a few key points. An antiferromagnetic order is stabilized on the Ni atomic sites with a magnetic moment of $1.516$ $\mu_{B}$. Breaking the magnetic moments into their orbital components we find the Ni $d_{x^2-y^2}$, $d_{z^2}$, and $t_{2g}$ have a moment of $0.7070$ $\mu_{B}$,  $0.7749$ $\mu_{B}$, and  $0.0364$ $\mu_{B}$, respectively. The apical oxygen atoms exhibit a $0.054$ $\mu_{B}$ moment collinear to the nickel atom at the center of the octahedron. The in-plane oxygen sites are polarized, but display no net moment. The shading of the octahedra in Fig.\ref{fig:BANDSanndSTRUCTURE}(c) follows the $(\pi,\pi)$ AFM ordering. 

The AFM phase opens a $1.64$ eV electronic band gap.~\footnote{Similar to Ref. \onlinecite{zhou2009synthesis}, we also found the high-temperature tetragonal crystal structure to produce a metal, due to $d_{x^2-y^2}$ and $d_{z^2}$ orbital overlap.}  The band gap develops in the half-filled $d_{x^2-y^2}$ and $d_{z^2}$ dominated bands by splitting the up- and down-spin antibonding level. A ``mirrored" splitting occurs around $-6.5$ eV in the bonding band, which breaks its spin degeneracy. The splitting at -6.5 eV occurs along the $\Gamma-M-\bar{\Gamma}$ cut in the Brillouin zone producing a gap of 0.5 eV. Interestingly, the electronic spectrum is not fully gaped out due to the presence metallic bands along $\bar{\Gamma}-\Gamma$ primarily of strong O $p_{x}+p_{y}$ character. Using the scheme presented in Ref.~\onlinecite{lane2018antiferromagnetic} the on-site Hubbard potental on the $d_{x^2-y^2}$ and $d_{z^2}$ orbitals is estimated to be $4.752$ eV and $5.481$ eV , respectively, along with a Hund's coupling of $0.519$ eV.

Since we are mainly interested in examining the excitonic behavior of La$_2$CuO$_4$ and La$_2$NiO$_4$, we concentrate our comparison of electronic structure to the states at the valence and conduction band edges. In LCO, the valence band is composed of $d_{x^2-y^2}$ and $d_{z^2}$  character bands. The relatively narrow $d_{z^2}$ bands are spread throughout the Brillouin zone, except for significant hybridization with $d_{x^2-y^2}$ around $M$ and a pure $d_{x^2-y^2}$ rising band along $\bar{\Gamma}-\Gamma$. The presence of $d_{z^2}$ character states at the valence edge is driven by Hund's coupling ($J_{H}\approx 1.248$ eV) present on the copper sites.~\cite{lane2018antiferromagnetic} The conduction band is highly dispersive and is composed of pure $d_{x^2-y^2}$ orbital character. In contrast, the valence in LNO is comprised of narrow intertwining $d_{x^2-y^2}$ and $d_{xy}$ character bands. The $d_{x^2-y^2}$ and $d_{xy}$ states do not appear to hybridize with one another, suggesting the valence is an even mixture of both orbitals.  The conduction band is essentially completely flat and of pure $d_{z^2}$ character, forcing any electron carriers to be extremely localized with a divergent effective mass. The offset in energy between $d_{x^2-y^2}$ and $d_{z^2}$ bands is driven by the tetragonal splitting of the $e_{g}$ levels and the presence of the $d_{xy}$ band is facilitated by Hund's coupling. We further emphasize that, due to the sizable $d_{z^2}$ and $d_{xy}$ contribution to the valence states in LCO and LNO, respectively, the conventional one-band model of the cuprates is of limited reach\cite{sakurai2011imaging}, as is the classification of the cuprates and nickelates within the Zaanen-Sawatzky-Allen\cite{zaanen1985band} scheme.

\begin{figure}[ht]
\includegraphics[width=0.99\columnwidth]{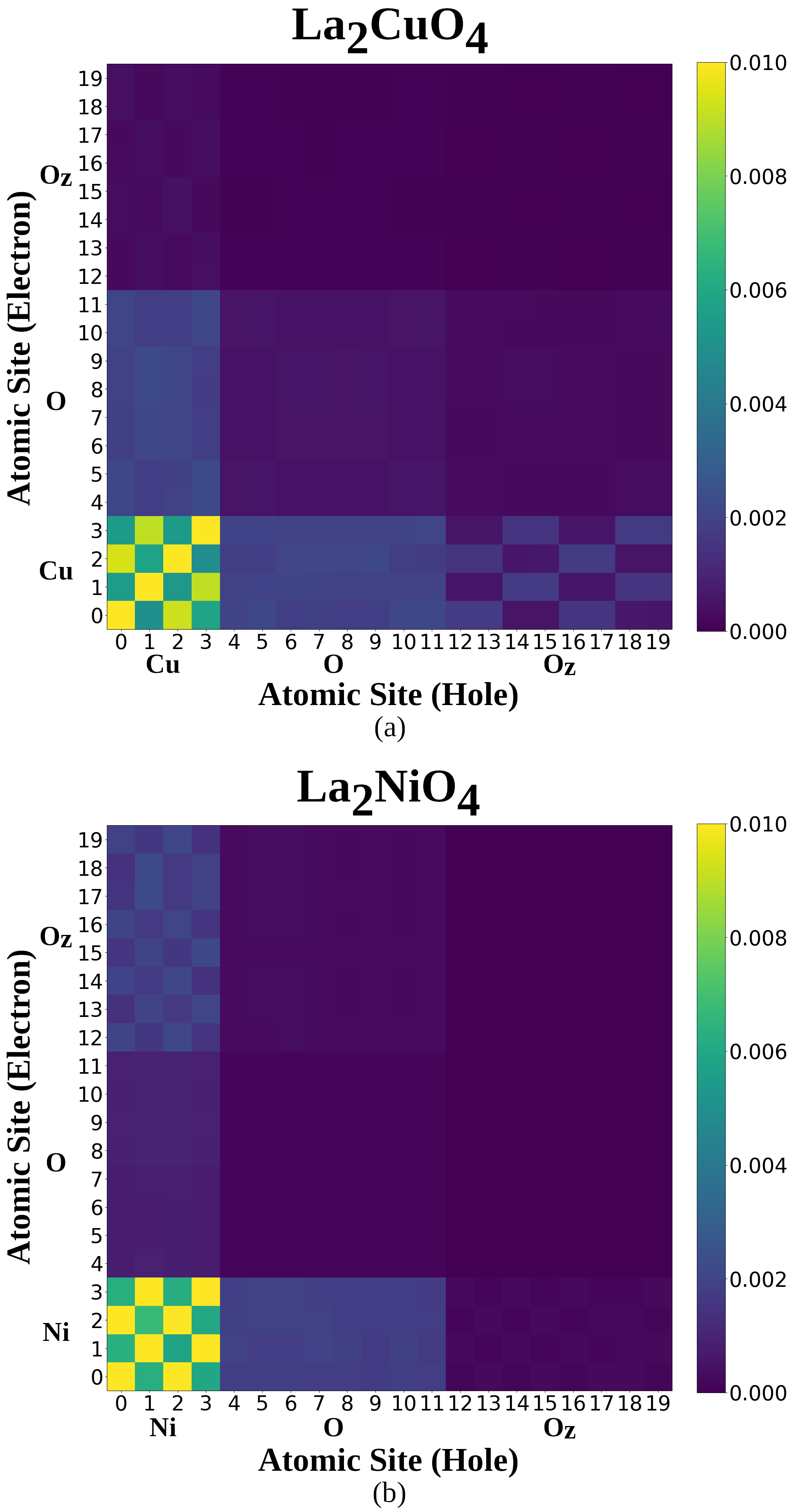}
\caption{(color online) Heat map of the atomic-site projected electron-hole coupling amplitude, $|C^{S}_{\tau, \tau^{\prime}}|^2$, for the lowest energy exciton in La$_2$CuO$_4$ and La$_2$NiO$_4$. The positions of the various atoms in the unit cell are given in Table~\ref{table:LCO} of  Appendix~\ref{a:pos}. } 
\label{fig:PROJEXC}
\end{figure}

\section{Excitonic Properties} 
Figure~\ref{fig:EXCBANDS}(a) and \ref{fig:EXCBANDS}(c) shows the energy dispersion of the first 100 excitonic states\footnote{We note the excitonic states analyzed in this work are not necessarily optically bright, but can be visible through the indirect RIXS scattering process.} along the high symmetry directions in the first Brillouin zone of La$_2$CuO$_4$ and La$_2$NiO$_4$, respectively. For LCO, there is a finite splitting between the lowest mode and the doubly degenerate pair of excitons $100$ meV higher in energy at $\Gamma$. These three excitons are quite dispersive throughout the Brillouin zone and are  separated in energy from the rest of the bands sitting at 0.7 eV and above.  Along the $\Gamma - X$ path,  the three lowest energy states follow a parabolic line shape in agreement with reported RIXS observations.\cite{collart2006localized,kim2002resonant} At the $X$ point,  the bands are separated by $20$ meV, but continuing along the $X - M$  path, the bands become nearly degenerate at $M$.  \footnote{The splitting between the excitonic modes follows the two band model proposed in Refs.~\onlinecite{matiks2009exciton,barford2002excitons}, where the next-nearest neighbor hopping parameter ($t^{\prime}$) is proportional to the splitting as $\delta E \approx 2t^{\prime}\cos(Q)$. Comparing to the splitting at $\Gamma$, $t^{\prime}\approx 36.5$ meV, which is very similar to tight-binding parameterizations giving $40$ meV.\cite{markiewicz2005one} }  Finally, the states are again split along $M - \Gamma$. Overall, the excitons in LCO appear to be quite mobile despite the background AFM order.

In LNO the excitonic states display very different behavior. Figure~\ref{fig:EXCBANDS} (c) shows a series of nearly flat bands, similar to atomic levels, with more dispersive bands staring at  $1.5$ eV. Each set of flat bands is triply degenerate, in contrast to the finite splitting found in LCO. The flat, non-dispersive nature of the excitonic states is in good accord with RIXS experiments.\cite{collart2006localized} The origin of flat bands is a direct consequence of the extremely localized $d_{z^2}$ band at the conduction edge of the ground state electronic structure, where the divergent effective mass of the electron effectively pins the exciton. \cite{cudazzo2015exciton}

To elucidate the band features and gain more insight into the localization of these lowest energy electron-hole pairs, we break down each exciton state into its component transitions. That is, the exciton wave function can be written as a linear combination of electron-hole pairs 
\begin{align}\label{eq:proj}
\ket{S^\mathbf{Q}}=\sum_{\mathbf{k}cv\sigma} Z^{S\sigma\mathbf{Q}}_{cv\mathbf{k}} \ket{cv\sigma \mathbf{k}\mathbf{Q}}
\end{align}
where $S$ indexes the excitonic state, $v$ ($c$) index the occupied (unoccupied) bands, $\sigma$ is the spin of the electron population, $\mathbf{k}$ ($\mathbf{Q}$) is the (center-of-mass) crystal momentum in the first Brillouin zone, and  $Z^{S\sigma\mathbf{Q}}_{cv\mathbf{k}}$ electron-hole amplitude, or equivalently an eigenvector of the excitonic Hamiltonian defined in Ref.~\onlinecite{sander2015beyond} which combines both resonant and antiresonant components of the 'super' vector $(A^S,B^S)$.

\begin{figure*}[ht]
\includegraphics[width=0.99\textwidth]{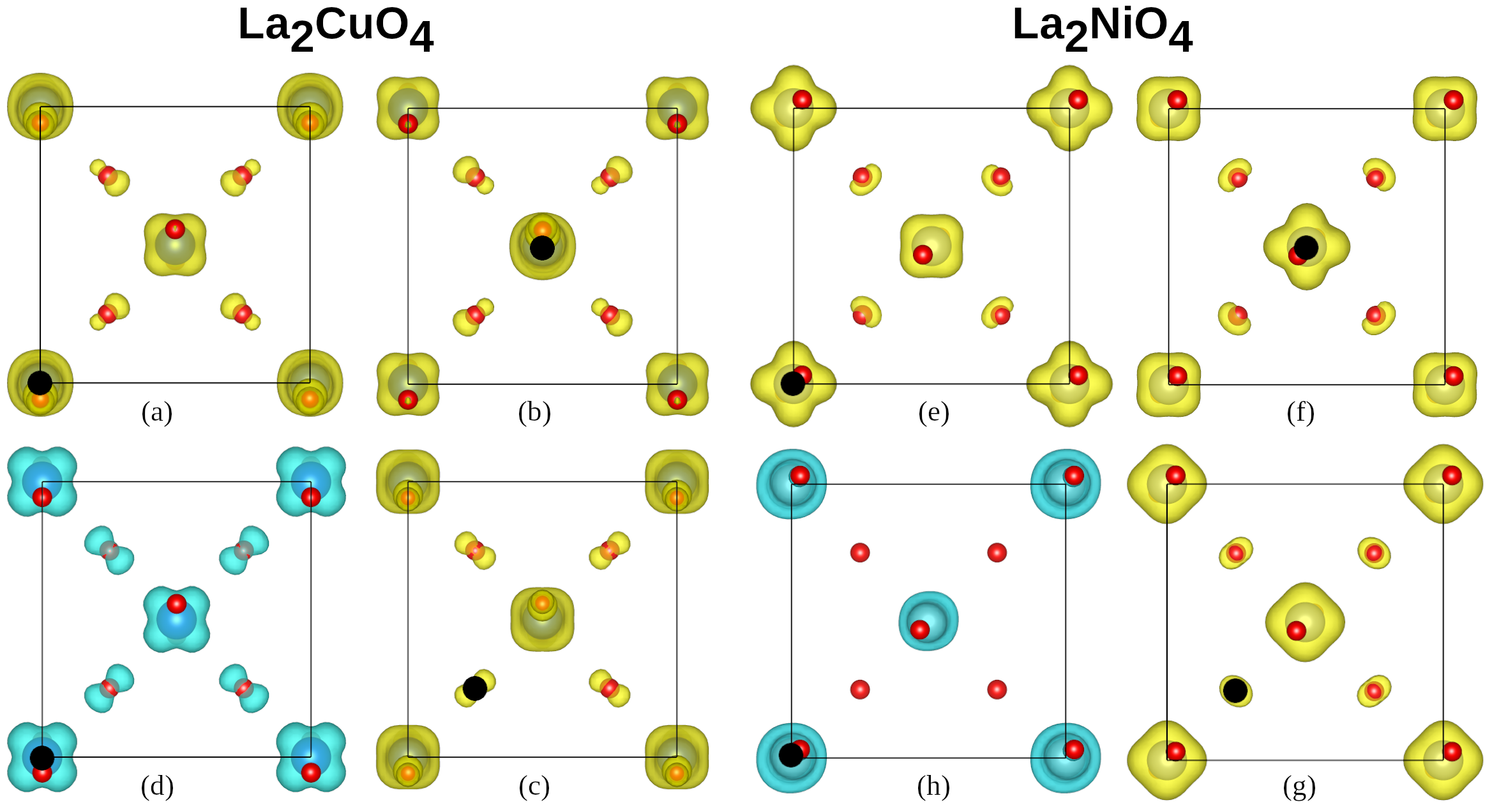}
\caption{(color online) The excitonic wave function of the lowest energy electron-hole pair in La$_2$CuO$_4$ (a)-(c) and La$_2$NiO$_4$ (e)-(g) when the electron is fixed at the black point. (d) and (h) shows the same for a fixed hole. The copper, nickel, and oxygen atoms are represented by blue, silver, and red spheres, respectively. } 
\label{fig:EXCWAVE}
\end{figure*}

Figure~\ref{fig:EXCBANDS}(b) and \ref{fig:EXCBANDS}(d) show the contribution of each momenta $\mathbf{k}$ in first Brillouin zone to the energetically lowest exciton at each high-symmetry $\mathbf{Q}$-point. The size and color of each dot goes as $\sum_{cv\sigma}  |Z^{S\sigma}_{cv\mathbf{k}}|^2$, where the contribution of each band (valence and conduction) and electron spin is integrated out. In LCO [Fig.~\ref{fig:EXCBANDS} (b)], the dominant crystal momentum of lowest energy exciton is centered on $(\pi,\pi)$ for $\mathbf{Q}=(0,0)$. For $\mathbf{Q}=(\pi,0)$, the momentum of the electron and hole is centered on $(\pi/9,\pi)$ with a slightly decreased spread as compared to $\mathbf{Q}=(0,0)$. Moreover when $\mathbf{Q}=(\pi,\pi)$, the momentum distribution of the electron-hole pairs is significantly spread out from the  $(0,0)$ center. Overall the excitons in LCO are tightly localized in $\mathbf{k}$-space, implying they are delocalized in real space on the order of charge-transfer or Wannier excitons. Therefore, they are quite mobile in the CuO$_2$ layer, consistent with their dispersion. The momentum distribution of exciton two and three are virtually identical to  the lowest energy exciton.

In contrast, the momentum distribution associated with the lowest energy exciton in LNO [Fig.~\ref{fig:EXCBANDS}(d)] is more uniformly spread throughout the Brillouin zone, admitting only slight peaks at  $(\pi/4,\pi/4)$, $(\pi,\pi/3)$, and $(\pi/3,\pi/3)$ for $\mathbf{Q}$ at $\Gamma$, $X$, and $M$, respectively. The momentum distribution can be thought of as two distributions superimposed: one a uniform background driven by the flat $d_{z^2}$ conduction band and another with slight inhomogeneities from the narrow valence bands. Due to the dominance of the uniform momentum distribution, these excitons are very localized in real space and effectively immobile, reflecting their dispersion.

Important to classifying excitons, and the phenomenology of a strongly correlated electron system in general, is to ask on which atomic sites the electrons and holes sit within the material system. To address this question, we project $Z^{S\sigma}_{cv\mathbf{k}}$ onto each atomic site including the full manifold of atomic orbitals corresponding to the particular atomic species. To project $Z^{S\sigma}_{cv\mathbf{k}}$, we define the change-of-basis transformation, $P$, between band space and the atomic-site-orbital space by writing the Kohn-Sham wave functions as a linear combination of a set of projected localized orbitals~\cite{schuler2018charge}
\begin{align}
\ket{\phi_{\mathbf{k}n\sigma}}&=\sum_{\tau lm}\ket{Y^{\tau}_{lm}}  \braket{Y^{\tau}_{lm}|\phi_{\mathbf{k}n\sigma}}=\sum_{\tau lm}  P^{\mathbf{k}n\sigma}_{\tau lm}   \ket{Y^{\tau}_{lm}} ,
\end{align}
where $\tau$ indexes the site, $lm$ specify the real spherical harmonic $Y_{lm}$, and $P^{\mathbf{k}n\sigma}_{\tau lm} $ is the transformation between bases. Further details of the local projections are given in Appendix~\ref{a:pos}. Substituting into Eq.~(\ref{eq:proj}),
\begin{align}
\ket{S^\mathbf{Q}}=\sum_{\substack{ \tau lm \\ \tau^{\prime} l^{\prime}m^{\prime} } } \left(  \sum_{\mathbf{k}cv\sigma} Z^{S\sigma\mathbf{Q}}_{cv\mathbf{k}}   P^{\mathbf{k+Q}nc\sigma}_{\tau lm}  P^{*~\mathbf{k}v\sigma}_{\tau^{\prime} l^{\prime}m^{\prime} }   \right)  \ket{Y^{\tau}_{lm}}\ket{Y^{\tau^{\prime}}_{l^{\prime}m^{\prime}}}^{*},
\end{align}
we arrive at the electron-hole amplitude in the atomic-site-orbital basis,
\begin{align}
C^{S\sigma\mathbf{Q}}_{\tau lm , \tau^{\prime} l^{\prime}m^{\prime}}= \sum_{\mathbf{k}cv\sigma} Z^{S\sigma\mathbf{Q}}_{cv\mathbf{k}}   P^{\mathbf{k+\mathbf{Q}}c\sigma}_{\tau lm}  P^{*~\mathbf{k}v\sigma}_{\tau^{\prime} l^{\prime}m^{\prime} }  .
\end{align}
Since we only find one active orbital per atomic site, we further simplify the discussion by integrated out the spin and orbital degrees of freedom, $C^{S\mathbf{Q}}_{\tau, \tau^{\prime}}$. Table \ref{table:pairingpathways} gives the relevant electron-hole pairing pathways with their corresponding orbital character.

Figure~\ref{fig:PROJEXC}(a) shows a heat map of $|C^{S\mathbf{Q}=0}_{\tau, \tau^{\prime}}|^2$ for the lowest energy exciton in La$_2$CuO$_4$, where the horizontal and vertical axes are the atomic sites of the holes and electrons, respectively. Here and thereafter, we will distinguish the in-plane oxygen atoms from the apical oxygen atoms as O and O$_z$, respectively. Firstly, we notice there is a clear asymmetry about the diagonal, indicating a difference in localization of the electron and hole. For example, the hole has a higher probability of sitting on the various apical oxygen sites as compared to the electron which displays relatively negligible weight on those atoms. Overall, the copper-copper sector exhibits the largest amplitude, with lesser weight on the in-plane and apical oxygen atoms. Within the copper-copper sector, the highest probability for exciton formation is along the diagonal. That is, the electrons and holes tend to coexist on the same copper atomic site. The next highly weighted pairing arrangement comes between copper atoms with the same magnetic polarization, but on different CuO$_2$ layers, suggesting  the existence of interlayer $d-d$ excitons in LCO. Lastly, excitons may form between copper atoms of differing magnetic polarization giving rise to Mott-Hubbard excitons. The non-zero weight in the Cu-O and Cu-O$_{z}$ sectors is due to the strong hybridization between copper and oxygen within the CuO$_2$ plane and generate charge-transfer electron-hole pairs.

Figure~\ref{fig:PROJEXC}(b) shows the same as (a) except for La$_2$NiO$_4$. In this case, the weight in all sectors is reduced or close to zero except for  the Ni-Ni, Ni-O, and O$_z$-Ni matrix elements, indicating the enhanced localization in LNO. The Ni-Ni zone is the highest weighted, displaying a clear two toned `checkerboard' pattern. Similar to LCO, electron-hole pairing between A-A and B-B magnetic sublattices is highly favored, whereas exciton formation within the NiO$_2$ plane between A and B sublattices is weak. Moreover, inter-layer and intra-layer excitons are found to be equally probable. Due to the orbital structure at the valence band edge, LNO does not exhibit any Mott-Hubbard type excitons. This difference is the result of tetragonal distortion of NiO$_6$ octahedra,  which buries the occupied $d_{z^2}$ orbital band below the Fermi level 
[Fig.~\ref{fig:BANDSanndSTRUCTURE}(b)], making it irrelevant in the low-energy physics. A variety of weak charge-transfer excitons are predicted facilitated by strong nickel-oxygen hybridization. Table~\ref{table:pairingpathways} summarizes the various types of excitons predicted along with their orbital character.

\begin{table}[h]
\centering
\begin{tabular}{c|c|c|c}
\multicolumn{4}{c}{La$_2$CuO$_4$}\\\hline
Sublattice & Hole             &Electron          & Type\\\hline\hline
A-A (B-B) & Cu $d_{z^2}$     & Cu $d_{x^2-y^2}$ & {\it d-d } \\\hline
A-B       & Cu $d_{x^2-y^2}$ & Cu $d_{x^2-y^2}$ & {\it Mott-Hubbard}\\\hline
-&O  $p_x+p_y$     & Cu $d_{x^2-y^2}$ & {\it Charge-Transfer}\\\hline
A-A (B-B)&O$_z$ $p_z$      & Cu $d_{x^2-y^2}$ & {\it Charge-Transfer} \\\hline
-&Cu $d_{z^2}$     & O $p_x+p_y$ & {\it Charge-Transfer} \\\hline
\multicolumn{4}{c}{}\\
\multicolumn{4}{c}{La$_2$NiO$_4$}\\\hline
Sublattice & Hole             &Electron          & Type\\\hline\hline
A-A (B-B)&Ni $d_{xy}$      & Ni $d_{z^2}$ & {\it d-d} \\\hline
A-B&Ni $d_{x^2-y^2}$ & Ni $d_{z^2}$ & {\it d-d} \\\hline
-&O  $p_x+p_y$     & Ni $d_{z^2}$ & {\it Charge-Transfer} \\\hline
A-B&Ni $d_{xy}$      & O$_z$ $p_x+p_y$  & {\it Charge-Transfer} \\\hline
A-A (B-B)&Ni $d_{xy}$      & O$_z$ $p_z$  & {\it Charge-Transfer} \\\hline
A-A (B-B)&Ni $d_{x^2-y^2}$ & O$_z$ $p_x+p_y$  & {\it Charge-Transfer} \\\hline
A-B&Ni $d_{x^2-y^2}$ & O$_z$ $p_z$  & {\it Charge-Transfer} \\\hline
\end{tabular}
\caption{Dominant electron-hole pairing channels in La$_2$CuO$_4$ and La$_2$NiO$_4$. Inter- and intra-layer pairing configuration are found for each exciton type.}\label{table:pairingpathways}
\end{table}

To gain further insight into the real space extension of the excitons in La$_2$CuO$_4$ and La$_2$NiO$_4$, we plot the excitonic wave function associated with the lowest energy exciton.  The  excitonic  wave  function  in real space is obtained by  projecting $\ket{S^\mathbf{Q}}$ [Eq.~\ref{eq:proj}] onto the spatial coordinates of the electron and hole, yielding 
\begin{align}
\Psi^{S\mathbf{Q}}(\mathbf{r}_e,\mathbf{r}_h)=\sum_{\mathbf{k}cv\sigma} Z^{S\sigma\mathbf{Q}}_{cv\mathbf{k}}   \phi_{c\mathbf{k+Q}}(\mathbf{r}_e)\phi^*_{v\mathbf{k}}(\mathbf{r}_h)  \;,
\end{align}
where $\mathbf{r}_e$ and $\mathbf{r}_h$ are the real-space electron and hole coordinates and $\phi$ is the SCAN-based Kohn-Sham wave functions. To  represent  the  six-coordinate function, we fix the hole (electron) position and we plot the resulting electron (hole) density, e.g., $|\Psi^{S}(\mathbf{r}_e,\mathbf{r}_h=\mathbf{R})|^2$. 

Figure~\ref{fig:EXCWAVE}(a)-(c)  shows the excitonic wave function within a CuO$_2$ plane for various fixed electron locations. In panel (a), the electron is fixed to the lower corner of the unit cell with the corresponding hole density concentrated on the copper and oxygen sites. The hole density at the corner site resembles a $d_{z^2}$ orbital, while at the center of the plane it is of $d_{x^2-y^2}$ character. If the electron is moved to the center atom [panel (b)] the orbital character switches between atomic sites. Moreover, if the electron is placed on a planar oxygen [panel (c)]  both copper atomic sites appear to be a linear combination of $d_{z^2}$ and $d_{x^2-y^2}$. The density surrounding the oxygen sites seems to be driven mainly by strong Cu-O hybridization. As a result, the in-plane oxygen atoms develop a hole density of $s\pm p_{x}$ ($s\pm p_{y}$) symmetry.  Moreover, halos of hole density are found surrounding the apical oxygen atoms. Finally, Figure~\ref{fig:EXCWAVE} (d) shows the corresponding electron density for a hole fixed to the corner of the unit cell. Here, the density is almost equivalent to the magnetic density obtained in Ref.~\onlinecite{lane2018antiferromagnetic}, where each copper atomic site has a $d_{x^2-y^2}$ orbital and the oxygen atoms are composed of pure $p_x$($p_y$) orbitals. The orbital character of electron and hole density directly follows the conduction and valence band characters of the ground state electronic structure.

Figure~\ref{fig:EXCWAVE}(e)-(h) is the same as (a)-(d)  except for the NiO$_2$ plane. The hole density exhibits the same behavior as LCO but displaying  $d_{x^2-y^2}$ and $d_{xy}$ orbital characters on the nickel sites. Panel (h) shows the electron density highly localized  to the nickel atoms, resembling a pure $d_{z^2}$ state, as expected from the band structure.

\section{Discussion}
The atomic-site-orbital resolved exciton coupling amplitude of LCO and LNO displays a rich landscape of excitonic pairing configurations, including local intra-atomic, semi-local intralayer and non-local interlayer excitons. These excitonic modes go beyond single or three-band models, which are limited to Mott or charge-transfer type excitations.\cite{zhang1998theory} Interestingly, we find two dominating fundamental types of electron-hole pairs: Mott-Hubbard and $d-d$. The Mott-Hubbard type consists of pairing between transition-metal sites of opposite magnetic polarization. In this scenario, an electron on site A is promoted to the empty conduction orbital of site B, described by
\begin{align}
&\ket{S_{Mott}} = \nonumber\\
 &\sum_{\mathbf{k} \sigma}  
C^{S\sigma}_{\mathbf{k}A \eta , B \eta}  \left[ c^{\dagger}_{\mathbf{k}A\sigma \eta}  c_{\mathbf{k}B\sigma \eta} + c.c.\right] \ket{AFM}\ket{O},  
\end{align}
where $\eta$ is the $d_{x^2-y^2}$ orbital and $\ket{AFM}\ket{O} $ represents the spin and orbital configuration of the ground state. This process is identical to doublon-holon production. Previous works \cite{rademaker2013exciton,wrobel2002excitons} have shown that doublon-holon binding is easier than hole-hole binding due to additoinal exchange processes and favors a $d$-wave state. Additionally, doublon-holon pairing has been recognized to produce a rich array of novel phases including an exciton checkerboard crystal and exciton superfluid phases.\cite{rademaker2013exciton} Recently, Imada and Suzuki have proposed a link between doublon-holon condensation and the psudogap phase. They also argue that 
high-temperature superconductivity can be driven by dipole attraction of the Mott-Hubbard excitons.~\cite{imada2019excitons} Therefore, the presence of intra- and inter-layer Mott-Hubbard excitons in LCO, though weaker than the $d-d$ excitions, suggests the existence of hidden excitonic phases and a possible excitonic origin of the pseudogap. 

It has been customarily thought that excitons within strongly correlated materials could not be localized at the same atomic site because of the large on-site Coulomb interaction. \cite{matiks2009exciton} However, we find strongly localized $d-d$ exciton formation to be highly favored.
In the $d-d$ channel, electron-hole pairs are formed on the same transition-metal sites or interlayer sites of equivalent magnetic polarization. In this process an electron makes an transition between two orbital levels, schematically given by,
\begin{align}
&\ket{S_{d-d}} = \nonumber \\
&\sum_{\mathbf{k} \sigma }   C^{S\sigma}_{\mathbf{k}A \eta , A \bar{\eta}} \left[  c^{\dagger}_{\mathbf{k}A \sigma \eta}c_{\mathbf{k}A \sigma \bar{\eta} } +c.c. \right]  \ket{AFM}\ket{O},
\end{align}
where $\eta ~(\bar{\eta})$ denotes the $d_{z^2}~ (d_{x^-y^2})$ orbital. Therefore, the $d-d$ exciton pairing is identical to orbiton creation. Orbital excitations have a rich history, complimented by their intimate connection to Jahn-Teller physics.\cite{khomskii2014transition,khomskii2010basic,kugel1982jahn} Theoretical studies of two-band models revealed an interesting interplay between spin and orbital degrees of freedom, producing orbital quasiparticles that propagate analogously to that of a hole in the 
AFM background.~\cite{van1998elementary,wohlfeld2011intrinsic,heverhagen2018spinon} Furthermore, spin and orbital interactions can promote bound states with a dispersion similar to the low-lying excitonic states found in Fig.~\ref{fig:EXCBANDS}. Signatures of these delicate new excitations have recently been reported on 
Sr$_2$CuO$_3$~\cite{schlappa2012spin} and Sr$_2$IrO$_4$.~\cite{kim2014excitonic} Within this picture, the difference in exciton dispersion between LCO and LNO can be interpreted as a sensitive balance between electron hopping, on-site repulsion, and Hund's coupling. The similarity of the excitonic dispersion and atomic-site-orbital breakdown, suggest a new  {\it ab initio} approach to modeling these exotic quasiparticles.  

At present, there remain many divergent views regarding the nature of both the normal and the superconducting states as well as the origin of the pairing mechanism in the high-Tc cuprates. Many proposals of pairing glues have been put forth, including spin-fluctuations,~\cite{scalapino1986d,scalapino2012common,scalapino1995case}
plasmons,~\cite{kresin1988layer,ishii1993acoustic,bill2003electronic} and excitons,\cite{ginzburg1970excitonic,allender1973model,weber1988cud,weber1989cu,jarrell1988charge,imada2019excitons} 
 each capturing various aspects of the cuprate phenomenology. However, the view that spin-fluctuations play a central role in determining the physical properties of the cuprates has been gaining increasing support. Complimenting the spin-fluctuations, we find nearly degenerate $d$ levels in both LCO and LNO at the valence band edge giving rise to a dominant $d-d$ exciton. This is indicative of strong low lying orbital excitations, which have been shown to strongly enhance spin-fluctuations. \cite{zaanen1991orbital} Suggesting possible synergistic cooperation between spin and orbital degrees of freedom could play a role in the anomalous nature of the cuprates.

Finally, we wish to comment on the classification of LCO and LNO as charge-transfer or Mott insulators. Within the Zaanen-Sawatzky-Allen\cite{zaanen1985band} scheme two competing energy scales are compared: the on-site Hubbard interaction $U$ and the charge-transfer energy $\Delta$. If $U\ll\Delta$, the lowest energy excitations are obtained by transferring one electron from one transition metal ion to anther --a Mott insulator--. On the other hand, if $U\gg\Delta$, the lowest energy excitations are from the ligand atoms to the transition metal --a charge-transfer insulator--.  However, our electronic structure shows a deviation from this scheme. Due to the significant presence of filled $d$-states at the valence band edge, under the influence of $J_{H}$, not $U$, the classification becomes ambiguous. As illustrated by the electron-hole pairing channels seen in  Fig.~\ref{fig:PROJEXC} and listed in Table~\ref{table:pairingpathways}, the position of the excited hole is diverse, exhibiting both Mott and charge-transfer behavior.  Interestingly, our results suggest that the electronic gap is predominately of $d-d$ type, where lowest energy excitations are obtained by transferring an electron from one orbital of the transition metal ion to anther. Further suggesting the presence of non-negligible orbital degrees of freedom.

\section{Concluding Remarks}\label{sec:conclusion}
In conclusion, our study demonstrates how excitonic excitations of complex correlated quantum materials can be addressed on a first-principles basis without the need to invoke ad hoc parameters or to restrict the orbitals included in the underlying Hamiltonian. Our finding of a myriad of different electron-hole pairing pathways, illustrates that the classification of correlated systems is more nuanced than the proposed Zaanen-Sawatzky-Allen criteria. Moreover, our study opens up a new pathway for examining the excited states of cuprates and other complex materials and their evolution with pressure and doping.


\begin{acknowledgments}
This work was carried out under the auspices of the U.S. Department of Energy (DOE) National Nuclear Security Administration under Contract No. 89233218CNA000001.
It was supported by the DOE Office of the Basic Energy Sciences Core Program (LANL Code: E3B5), and in part by the Center for Integrated Nanotechnologies, a DOE BES user facility, in partnership with the LANL Institutional Computing Program for computational resources.  
\end{acknowledgments}

\appendix

\section{Local Projection Details}\label{a:pos}
On each site a full set of real hydrogen-like functions $s$, $p$, and $d$ were employed using the default main quantum number of the hydrogen radial function. Details of the sites on which the local projections defined in Eq.~(\ref{eq:proj}) are centered within the crystal structure of LTO La$_2$CuO$_4$ and LTT La$_2$NiO$_4$ are given in Table \ref{table:LCO}.

\begin{table}[h]
\centering
\begin{tabular}{l|c|c|c||c||l|c|c|c}
\hline\hline 
La$_2$CuO$_4$      & x    & y      & z          &&  La$_2$NiO$_4$          & x    & y         & z        \\\hline
Cu (0)& 0    & 0      & 0          && Ni (0)     & 0    & 0         & 0        \\
Cu (1)& 0    & 0.5    & 0.5        && Ni (1)     & 0.5  & 0         & 0.5      \\
Cu (2)& 0.5  & 0      & 0.5        && Ni (2)     & 0    & 0.5       & 0.5      \\
Cu (3)& 0.5  & 0.5    & 0          && Ni (3)     & 0.5  & 0.5       & 0        \\
O (4) & 0.25 & 0.25   & 0.011      && O (4)      & 0.25    & 0.25   & 0.989    \\
O (5) & 0.75 & 0.75   & 0.989      && O (5)      & 0.25    & 0.25   & 0.485    \\
O (6)      & 0.75 & 0.75   & 0.489 && O (6)      & 0.75    & 0.75   & 0.516    \\
O (7)      & 0.25 & 0.25   & 0.511 && O (7)      & 0.75    & 0.75   & 0.016    \\
O (8)      & 0.75 & 0.25   & 0.511 && O (8)      & 0.25    & 0.75   & 0        \\
O (9)      & 0.25 & 0.75   & 0.489 && O (9)      & 0.75    & 0.25   & 0.5      \\
O (10)     & 0.25 & 0.75   & 0.989 && O (10)     & 0.75    & 0.25   & 0        \\
O (11)     & 0.75 & 0.25   & 0.011 && O (11)     & 0.25    & 0.75   & 0.5      \\
O$_z$ (12) & 0    & 0.944  & 0.186 && O$_z$ (12) & 0.031   & 0.031  & 0.117    \\
O$_z$ (13) & 0    & 0.056  & 0.814 && O$_z$ (13) & 0.469   & 0.031  & 0.677    \\
O$_z$ (14) & 0    & 0.444  & 0.314 && O$_z$ (14) & 0.031   & 0.469  & 0.677    \\
O$_z$ (15) & 0    & 0.556  & 0.686 && O$_z$ (15) & 0.531   & 0.969  & 0.323    \\
O$_z$ (16) & 0.5  & 0.944  & 0.686 && O$_z$ (16) & 0.969   & 0.531  & 0.323    \\
O$_z$ (17) & 0.5  & 0.056  & 0.314 && O$_z$ (17) & 0.469   & 0.469  & 0.177    \\
O$_z$ (18) & 0.5  & 0.444  & 0.814 && O$_z$ (18) & 0.531   & 0.531  & 0.823    \\
O$_z$ (19) & 0.5  & 0.556  & 0.186 && O$_z$ (19) & 0.969   & 0.969  & 0.823    \\
\hline\hline
\end{tabular}
\caption{\label{table:LCO}
The sites on which the local projections are centered within the crystal structure of LTO La$_2$CuO$_4$ and LTT La$_2$NiO$_4$ in units of the lattice vectors.}
\end{table}

\bibliography{Refs}

\end{document}